\pdfoutput=1 
\documentclass[11pt,twocolumn,bahasa,canadian,english,titlepage]{IEEEtran}
\usepackage[T1]{fontenc}
\usepackage[latin9]{inputenc}
\usepackage[a4paper]{geometry}
\usepackage[active]{srcltx}
\usepackage{float}
\usepackage{amsthm}
\usepackage{amsmath}
\usepackage{amssymb}
\usepackage{graphicx}
\usepackage{perpage}
\usepackage{fancyhdr}
\usepackage{graphicx}
\usepackage{multicol}
\usepackage{lettrine}
\usepackage{icomma}
\usepackage{microtype}
\usepackage{algorithm2e}
\MakePerPage{footnote}

\makeatletter

\floatstyle{ruled}
\newfloat{algorithm}{tbp}{loa}
\providecommand{\algorithmname}{Algorithm}
\floatname{algorithm}{\protect\algorithmname}

\theoremstyle{plain}
\newtheorem{thm}{\protect\theoremname}
\theoremstyle{definition}
\newtheorem{defn}[thm]{\protect\definitionname}
\theoremstyle{remark}
\newtheorem{rem}[thm]{\protect\remarkname}
\theoremstyle{plain}
\newtheorem{prop}[thm]{\protect\propositionname}
\theoremstyle{definition}
\newtheorem{example}[thm]{\protect\examplename}

\@ifundefined{date}{}{\date{}}

\makeatother

\usepackage{babel}
\addto\captionsbahasa{\renewcommand{\algorithmname}{Algoritma}}
\addto\captionsbahasa{\renewcommand{\definitionname}{Definisi}}
\addto\captionsbahasa{\renewcommand{\examplename}{Contoh}}
\addto\captionsbahasa{\renewcommand{\propositionname}{Proposisi}}
\addto\captionsbahasa{\renewcommand{\remarkname}{Catatan}}
\addto\captionsbahasa{\renewcommand{\theoremname}{Teorema}}
\addto\captionscanadian{\renewcommand{\algorithmname}{Algorithm}}
\addto\captionscanadian{\renewcommand{\definitionname}{Definition}}
\addto\captionscanadian{\renewcommand{\examplename}{Example}}
\addto\captionscanadian{\renewcommand{\propositionname}{Proposition}}
\addto\captionscanadian{\renewcommand{\remarkname}{Remark}}
\addto\captionscanadian{\renewcommand{\theoremname}{Theorem}}
\addto\captionsenglish{\renewcommand{\definitionname}{Definition}}
\addto\captionsenglish{\renewcommand{\examplename}{Example}}
\addto\captionsenglish{\renewcommand{\propositionname}{Proposition}}
\addto\captionsenglish{\renewcommand{\remarkname}{Remark}}
\addto\captionsenglish{\renewcommand{\theoremname}{Theorem}}
\providecommand{\definitionname}{Definition}
\providecommand{\examplename}{Example}
\providecommand{\propositionname}{Proposition}
\providecommand{\remarkname}{Remark}
\providecommand{\theoremname}{Theorem}

\begin{document}

\title{Graph Decompositions Analysis and Comparison for Cohesive Subgraphs
Detection\bigskip{}
}

\author{\begin{multicols}{2} Etienne \textsc{Callies}\\X2011 - École
  Polytechnique\\etienne.callies@polytechnique.edu\\Tomás \textsc{Yany
  Anich}\\X2011 - École Polytechnique\\tomas.yany@polytechnique.edu
\end{multicols}
\noindent\rule{15cm}{0.3pt}}
\maketitle

\begin{abstract}
Massive networks have shown that the determination of dense subgraphs, where
vertices interact a lot, is necessary in order to visualize groups of common
interest, and therefore be able to decompose a big graph into smaller
structures. Many decompositions have been built over the years as part of
research in the graph mining field, and the topic is becoming a trend in the
last decade because of the increasing size of social networks and databases.
Here, we analyse some of the decompositions methods and also present a novel
one, the Vertex Triangle \textit{k}-core. We then compare them and test them
against each other. Moreover, we establish different kind of measures for
comparing the accuracy of the decomposition methods. We apply these
decompositions to real world graphs, like the \emph{Collaboration network of
arXiv} graph, and found some interesting results.
\end{abstract}


\section{Introduction}

Graphs representing real data are nowadays evolving at
a speed we can hardly control. They are bigger than before, they grow
up faster and they are more complex as well. The information boom
in the last decade has been a crucial motivation for understanding
and analysing in a more efficient way large graphs. Social networks,
for example, are represented with graphs that can go up to billions
of vertices, and more. Therefore, it is compulsory to build new tools
that would allow us to identify smaller structures, denser subgraphs
or at least more connected vertices that could represent communities
inside this big pool: the graph itself.

In fact, clustering is an important problem in graph mining. How to
determine if there are, and which are they, groups inside a large
graph? Learning this represent a considerable issue in two senses:
how to identify and calculate in an effective and efficient way this
dense subgraphs, and how to determine if the results are accurate
as we would expect, i.e. if they in fact reveal real communities.
Once we have answered this, we can no longer preoccupy ourselves with
the big and complex graph, but more with its smaller decompositions
which represent better and denser communities. This could be useful,
for instance, for targeted advertising, news clustering and others.

Many methods, that we will review, have been presented over the last
few years. We no longer found graph's description in the vertices'
degree\footnote{It can be shown that high degree nodes can coexist with low degree
nodes.}. Today, we need better tools that could make the difference between
kinds of communities --if we are talking about social graphs-- or
clusters, in the general case.

In this paper we will review some of the principal modern decompositions
techniques, analyse their algorithms, compare them and evaluate them.
The computing time will be important in our task since we need algorithms
not only to be good, but also to be efficient. Our contributions lie
on the regrouping and side by side comparison between existing, and
an innovative, decomposition techniques, and on a novel metric for
evaluation of the detected important subgraphs.

We will be working with a simple undirected graph $G=(V,E)$ where
$V$ is the set of vertices and $E$ the set of edges. Moreover, we
will consider the degree $d_{G}(v)$ of a vertex $v\in V$ to be the
number of edges $e\in E$ incident to the vertex itself. Some others
assumptions and definitions will be explained when needed.

\section{Graph Decompositions }

We would begin with some important definitions that, although are
typical among the computer science community, are always necessary
to clarify.
\begin{defn}
A graph $G$ is said to be \emph{complete} when every of its vertices
is connected to each other, i.e., $\forall(x,y)\in V^{2},x\neq y,\exists e\in E$
with $e=\{x,y\}$.
\end{defn}
When we are thinking of finding the most connected subgraph inside
a graph, we are somehow thinking on the \emph{clique problem}, i.e.,
finding cliques in a graph $G$. The definition of a \emph{clique}
might in fact vary according to the literature, so we prefer to establish
the one we think it is the most used.
\begin{defn}
Let $G'\subset G$ be a subgraph of a graph $G$. We say that $G'$
is a \emph{clique} if it is a complete subgraph of $G$. We say that
$G'$ is a \emph{clique of degree $k$} if it is a clique where each
vertex has degree $k$
\end{defn}
Our quest should be then to find maximal size \emph{cliques} in a
graph $G$, in a way, to solve the \emph{clique problem}. This presents
two majors issues, as said by \cite{key-1}, cliques of small size
are too numerous for having any kind of special interest, and those
cliques which are big enough for being interesting are too rare to
find. The second problem is the complexity of the clique problem:
listing every possible size clique --if there are any-- can be computed
in exponential time.

Therefore, finding other kind of interesting subgraphs of a graph,
which could represent dense and almost complete structure, without
being a clique, appear to be an important task, and a useful one,
especially if we can show that this task can be computed in polynomial
time. We will here review some of the most known decomposition one
can find in literature.

\begin{figure}
\begin{centering}
\includegraphics{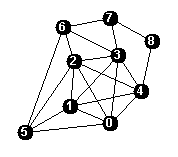}
\par\end{centering}

\protect\caption{Simple undirected graph $G_{0}$\label{fig:Simple-undirected-graph}}

\end{figure}

We will use Figure \ref{fig:Simple-undirected-graph} as our canonic
example through the whole paper.

\subsection{$k$-core}

\begin{figure}
\begin{centering}
\includegraphics{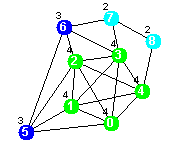}
\par\end{centering}

\protect\caption{$k$-core decomposition of $G_{0}$\label{fig:-core-decomposition-of}}
\end{figure}

As defined in \cite{key-2}, we use the following notations.
\begin{defn}
Let $G$ be a simple undirected graph. We define $\Delta(G)=\min_{v\in V}d_{G}(v)$,
the minimum degree of a vertex in $G$.
\end{defn}
Having this definition set, we can define what a $k$-core is.
\begin{defn}
Let $G'\subset G$ be a subgraph of a graph $G$. We say that $G'$
is the \emph{$k$-core} of $G$ if it is the maximum size subgraph
of $G$ where $\Delta(G')\geq k$.
\end{defn}
We can immediately notice that this approach is more relaxed than
the clique's one. In fact, $k$-cores are restrained subgraphs, but
not as much as cliques. We ask them to be as connected as we want
them to --varying the $k$-- but we don't ask them to be \emph{the
most connected}. Nevertheless, $k$-cores represent very connected
structures, as said in \cite{key-3}, they are \textquotedbl{}seedbeds,
within which cohesive subsets can precipitate out\textquotedbl{}.
Therefore their study become important as we will see in the following
parts of this paper. Moreover, some interesting analysis can be done
for each vertex of a graph when a $k$-core decomposition is made.
\begin{defn}
Let $v\in V$ be a vertex of a graph $G$. The \emph{core number}
of $v$ is the highest $k$ for which $v$ belongs to the $k$-core
of $G$.
\end{defn}
This last definition will be very important in our examples because
it will allow us to considerate what we will call\emph{ levels }in
a decomposition.
\begin{defn}
Given a graph decomposition, we call \emph{level} the set of vertices
having the same core number.
\end{defn}
Figure \ref{fig:-core-decomposition-of} shows a $k$-core decomposition
over the initial graph $G_{0}$. The numbers next to the vertices
represent the \emph{core number}. Different colours here represent
different levels.

\begin{rem}
Some interesting assets can be detected in $k$-cores. In fact, as
shown in \cite{key-4}, $k$-cores have the following properties:\end{rem}
\begin{itemize}
\item $k$-cores are nested, i.e., let $k<k'$ and $G,G'$ $k$-core and
k'-core of a graph $G$ $\Rightarrow G'\subseteq G$.
\item $k$-cores can have more than one connected component.\end{itemize}
\begin{prop}
If we consider a connected $k$-core $G'=(V',E')$ of a graph $G=(V,E)$
and we define $n=\big|\{v,v\in V'\}\big|$ then, if $k=n-1$, $G'$
is a clique. \end{prop}
\begin{IEEEproof}
If $k=n-1$ then each vertex on the $k$-core is connected to all
of its neighbours. This is the definition of a clique.\end{IEEEproof}
\begin{rem}
It is clear that if the $k$-core has more than one connected component
then the result can be apply to each one of them.
\end{rem}
This last proposition show us that a $k$-core is very close to a
clique, without being specifically a clique itself. Knowing this,
we will develop in the following parts of this paper some measures
for the difference between a deduced subgraph and a clique, and see
how close were we to find a proper clique. Actually, with the definition
of a clique of degree $k$ given before, it is trivial that, if such
clique of degree $k$ exists in a graph $G$, then it will be a subset
of a $k$-core $G'$ of $G$.

\subsection{Generalized cores}

The idea behind $k$-cores can be extended. When we talked about $k$-cores
we were considering the function $\Delta(G)$ as the minimum degree
of a vertex in a graph $G$. What if we consider another different
function to which we ask some properties? This generalization as seen
in \cite{key-4} can allow us to define more general decompositions
than $k$-cores.
\begin{defn}
Let $Z=\{z:z\subseteq V\}$. Let $p:V\times Z\rightarrow\mathbb{R}$
be a function. We say that $p$ is a \emph{vertex property function},
or a $p$-\emph{function.}
\end{defn}
We can now define the notion of $p$-core at level $t$, which come
to extend the idea of $k$-core.
\begin{defn}
Let $G_{p}^{t}=(V_{p}^{t},E_{p}^{t})$ be a subgraph of a graph $G$.
We say that $G_{p}^{t}$ is a $p$\emph{-core at level $t\in\mathbb{R}$}
if and only if \end{defn}
\begin{itemize}
\item $\forall v\in V_{p}^{t}:t\leq p(v,V_{p}^{t})$
\item $V_{p}^{t}$ is a maximal subset with this property\end{itemize}
\begin{rem}
We can notice that the function used for the $k$-core decomposition
is in fact a particular $p$-function, where for $v\in V,U\subseteq V$
we have $p(v,U)=\min_{v\in U}d_{U}(v)$.
\end{rem}
As said before, we need to ask some particular properties to this
$p$ function so it can be coherent with what we expect from decompositions.
\begin{defn}
Let $p$ be a vertex property function. We say that $p$ is \emph{monotone
}if and only if for $U\subset U'\subset V$ and $\forall v\in U$
we have $p(v,U)\leq p(v,U'$).
\end{defn}
This definition allow us to enunciate an important property for decompositions.
\begin{prop}
\label{monotony}Let $p$ be a monotone function. $p$-cores are nested,
i.e., for $t_{2}<t_{1}\Rightarrow V_{p}^{t_{1}}\subseteq V_{p}^{t_{2}}$
where $V_{p}^{t_{1}}$ and $V_{p}^{_{2}}$ are the vertices of the
$p$-core at level $t_{1}$ and $t_{2}$ respectively.\end{prop}
\begin{IEEEproof}
Let $t_{2}<t_{1},v\in V_{p}^{t_{1}}$. By definition we have $t_{1}\leq p(v,V_{p}^{t_{1}})$.
Besides, $t_{2}<t_{1}.$ Therefore, $t_{2}<p(v,V_{p}^{t_{1}})$. But,
by definition, $V_{p}^{t_{2}}$ is the maximal subset $U$ such that
$t_{2}\leq p(v,U)$. Therefore $V_{p}^{t_{1}}\subseteq V_{p}^{t_{2}}$.\end{IEEEproof}
\begin{example}
Many functions can be chosen for constructing $p$-core at level $t$
graphs. \end{example}
\begin{itemize}
\item $p_{1}(v,V')=d_{G'}(v)$. With this function one can get the $k$-cores
defined before.
\item $p_{2}(v,V')=$ number of different cycles of length $\ensuremath{3}$
(triangles) going through $\ensuremath{v}$. We will analyse this
particular function in a future part of this paper.
\item For a directed graph $G$ we can consider the following $p$-functions:
$p_{3}(v,V')=d_{G'}^{in}(v)$ and $p_{4}(v,V')=d_{G'}^{out}(v)$.
\end{itemize}

\subsection{Triangle $k$-core and k-truss}

\begin{figure}
\begin{centering}
  \includegraphics{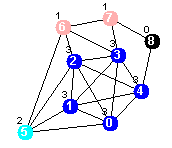}
\par\end{centering}

\protect\caption{Triangle $k$-core decomposition of $G_{0}$\label{fig:-truss-decomposition-of}}
\end{figure}

Triangle $k$-cores and $k$-trusses --which we will show are the
same-- are very important decompositions for our analysis. In fact,
latest studies \cite{key-5,key-6} have shown that in order to look
up for connected subgraphs inside a graph we need to search for vertices
that not only are connected between them, but that have common neighbours.
This last feature is actually represented through triangle $k$-cores
and $k$-trusses.
\begin{defn}
We call a \emph{triangle} a cycle of length $3$ inside a graph. Let
$G'$ be a subgraph of a graph $G$. We say that $G'$ is \emph{triangle
$k$-core }of $G$ if each edge of $G'$ is contained within at least
$k$ triangles in $G'$ itself.
\end{defn}
We define as well what a $k$-truss is.
\begin{defn}
Let $G'$ be a subgraph of a graph \textbf{$G$}. We say that $G'$
is a $k$\emph{-truss} of $G$ if each edge of $G'$ connects two
vertices who have at least $k-2$ common \foreignlanguage{bahasa}{neighbours}.
\end{defn}
Figure \ref{fig:-truss-decomposition-of} shows the triangle $k$-core
decomposition in our graph $G_{0}$.

\begin{prop}
A $k$-truss is a triangle ($k-2$)-core.\end{prop}
\begin{IEEEproof}
In a $k$-truss, in order to have $k-2$ common \foreignlanguage{canadian}{neighbours},
every edge must be reinforced by at least $k-2$ pairs of edges: they
form therefore $k-2$ triangles.
\end{IEEEproof}
Although some literature has been written for one and the other kind
of decomposition, we have just shown that they get the same result
at the end. Because of this, we will be working the whole time with
the \emph{triangle $k$-core }term.
\begin{prop}
Triangle $k$-core are nested subgraphs, i.e.,
$k_{2}<k_{1}\Rightarrow$ triangle $k_{1}$-core $\subseteq$ triangle
$k_{2}$-core.\end{prop}
\begin{IEEEproof}
Let $e\in E_{k_{1}}$, i.e. $e$ is contained within $k_{1}$ triangles.
Since $k_{2}<k_{1}$, $e$ is also contained in $k_{2}$ triangles.
So $e\in E_{k_{2}}$. \end{IEEEproof}
\begin{prop}
If $G$ has a clique of degree $k>1$ then it is contained within
a $(k+1)$-truss, i.e. within a triangle $(k-1)$\textup{\emph{-core}}. \end{prop}
\begin{IEEEproof}
Let $C=(V_{C},E_{C})$ be a clique of degree $k>1$ and $K=(V_{k},E_{k})$
a ($k+1$)-truss within a graph $G=(V,E)$. Since each vertex in $C$
has degree $k$, then $|V_{C}|=k+1$. Therefore, for each $e\in E_{C}$,
$e=\{u,v\}$, $u$ and $v$ will be \foreignlanguage{canadian}{neighbours}
with every other vertex in $C$, in particular, they will share $(k+1)-2$
\foreignlanguage{canadian}{common neighbours}. With this we have $E_{C}\subseteq E_{k}$.
Since a graph can be reduced to its edges set, when isolated vertices
don't exist, we have just shown that $C\subseteq K$.
\end{IEEEproof}

\subsection{Vertex triangle $k$-core}

\begin{figure}
\begin{centering}
\includegraphics{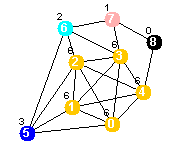}
\par\end{centering}

\protect\caption{Vertex triangle $k$-core decomposition of $G_{0}$\label{fig:Vertex-triangle--core}}
\end{figure}

We wanted to find another decomposition following the philosophy behind
the triangle $k$-core decomposition. Therefore, instead of considering
triangles around an edge, we will look for triangles around a vertex.
With this, we have the same idea as mentioned before: we try to identify
dense subgraphs in terms of vertices having common neighbours.
\begin{defn}
Let $G'$ be a subgraph of a graph \textbf{$G$}. We say that $G'$
is a \emph{vertex triangle $k$-core} if every vertex in $G'$ is
contained in at least $k$ triangles.\end{defn}
\begin{prop}
Vertex triangle $k$-core are nested subgraphs.\end{prop}
\begin{IEEEproof}
Using the proposition \ref{monotony} we just have to show that $p(v,U)=$
number of triangles through $v$, is a monotone function.

Let $U_{1}\subseteq U_{2}\subseteq V$. Since $U_{2}$ has more vertices
than $U_{1}$ we have two options: whether this new vertices are creating
triangles going through $v$ and therefore $p(v,U_{1})<p(v,U_{2})$,
whether they are not creating any new triangle going through $v$
and therefore $p(v,U_{1})=p(v,U_{2})$. In any case, we have $p(v,U_{1})\leq p(v,U_{2})$.
\end{IEEEproof}
Figure \ref{fig:Vertex-triangle--core} shows the vertex triangle
$k$-core decomposition of $G_{0}$. We notice immediately that this
decomposition is different from the (edge) triangle $k$-core decomposition.
\begin{rem}
The maximum core number for vertex triangle decomposition is $\frac{(n-1)(n-2)}{2}$
whereas for the (edge) triangle decomposition it goes up to $n-2$.
Moreover, for the $k$-core decomposition it goes only up to $n-1$.
\end{rem}

\section{Algorithms}

Once we have described the possible decompositions one can apply to
a graph, we would like to review the principal algorithms that are
associated with these subgraphs detection. We chose to do it in a
separate part because although definitions and properties are always
valid, algorithms can change during time and can in particular be
improved (in the best scenario). We will review algorithms for $k$-core,
triangle $k$-core and vertex triangle $k$-core. For any further
analysis we could make, these three decomposition are the only ones
to think of.

The general procedure for each algorithm will be the same. In fact,
since we will be asking to vertices --or edges-- to have some properties,
the method will almost always be to delete those vertices or edges
that do not fulfil the condition we are asking. Repeating this process
will get us to the final subgraph we are looking for. But the idea
for us will be to actually memorize the core number for each vertex,
therefore our result will not be a subgraph but a numbering over the
graph vertices.

In our study we consider the complexity as the evolution of the number
of elementary operations in relation to the number of vertices in
the input graph $G=(V,E)$. The complexity is said quadratic (resp.
cubic) if the number of operations grows like $n^{2}$ (resp. $n^{3}$)
where $n=|V|$ is the number of vertices.

\subsection{$k$-core }

One of the great advantages of $k$-core decomposition is it easy
way to compute it. The idea of the algorithm will be to take a graph
$G$ and delete recursively all vertices of degree less than $k$.
With them, we delete as well every edge incident with the concerned
vertices.

\begin{algorithm}
  \ForAll{$v \in V$}{
    $d(v) \leftarrow$ \# of edges that contain $v$\;
  }
  $k \leftarrow0$\;

  \While{$V\neq\emptyset$}{
  $k \leftarrow k+1$\;\\
  $unprocessedVertices = copy(V)$\;

  \While{$unprocessedVertices\neq\emptyset$}{
  Vertex $v = pop(unprocessedVertices)$\;

  \If{$d(v)<k$}{
    \ForAll{neighbours $w$ of $v$}
    {
    $d(w) \leftarrow d(w)-1$\;\\
    $unprocessedVertices.add(w)$\;
    }
    $v$.coreNumber$ \leftarrow k-1$\;\\
    remove $v$ from $V$\;
  }
  }
  }
\caption{$k$-core decomposition algorithm}
\end{algorithm}

In order to evaluate the complexity of the algorithm, we will first
consider the initialization, then the emptying of \emph{unprocessedVertices}
which contain vertices unprocessed by default, and also neighbours
vertices added after a remove.

\textbf{Initialization.} The calculation of degree is quadratic in
terms of vertices, because it is linear in terms of edges.

\textbf{Copied in }\textbf{\emph{unprocessedVertices}}\textbf{.} The
$k$ number goes from $0$ to $n-1$. There are therefore in all no
more than $n^{2}$ vertices copied in \emph{unprocessedVertices}.

\textbf{Unprocessed after a remove.} When a vertex is removed, all
its neighbours are added to \emph{unprocessedVertices}. There are
therefore in total no more than $n^{2}$ vertices added to \emph{unprocessedVertices}
after a remove.

\textbf{\emph{$\Rightarrow$}}\textbf{ Global complexity: }$\mathcal{O}(n^{2})$

\subsection{Triangle $k$-core}

Triangle $k$-cores are a bit more difficult to compute than $k$-cores
since we have to considerate triangles through edges. Nevertheless,
the general form of the algorithm remains similar.

\begin{algorithm}
  \ForAll{$v \in V$}{
    $d(v) \leftarrow$ \# of edges that contain $v$\;
  }

  \ForAll{$e \in E$}{
    $triangle(e) \leftarrow$ \# of triangles that contain $e$\;
  }

  $k \leftarrow0$\;

  \While{$E\neq\emptyset$}{
  $k \leftarrow k+1$\;\\
  $unprocessedEdges = copy(E)$\;

  \While{$unprocessedEdges\neq\emptyset$}{
  Edge $e = pop(unprocessedEdges)$\;

  \If{$triangle(e)<k$}{
    \ForAll{triangle $t$ of $e$}
    {
      \For{both other edges $f$ of $t$}
      {
        $triangle(f)\leftarrow triangle(f)-1$\;\\
        $unprocessedEdges.add(f)$\;
      }
    }
    \For{both vertices $v$ of $e$}
    {
      $d(v)\leftarrow d(v)-1$\;\\
      \If{$d(v)=0$}
      {
        $v$.coreNumber$\leftarrow k-1$\;
      }
      remove $e$ from $E$\;
    }
  }
}
}
\caption{Triangle $k$-core decomposition algorithm}
\end{algorithm}

\textbf{Initialization.} The calculation of degree is quadratic, and
the calculation of \emph{triangle($f$)} for an edge $f$ is linear,
so globally cubic complexity.

\textbf{Copied in }\textbf{\emph{unprocessedEdges}}\textbf{.}\emph{
}The $k$ number goes from $0$ to $n-2$. There are therefore in
total no more than $n^{3}$ edges copied in \emph{unprocessedEdges}.

\textbf{Unprocessed after a remove.} When an edge is removed, all
other edges which formed a triangle with the removed edge are added
to \emph{unprocessedEdges}. The number of such edges is not more than
twice the number of vertices. There are therefore in total no more
than $n^{3}$ edges added to \emph{unprocessedEdges} after a remove.

\textbf{\emph{$\Rightarrow$}}\textbf{ Global complexity:}\emph{ }$\mathcal{O}(n^{3})$

\subsection{Vertex triangle $k$-core}

Vertex triangle $k$-core are almost as difficult to compute as (edge)
triangle $k$-core, with the advantage we only go through vertices
and never look for edges.

\begin{algorithm}
  \ForAll{$v \in V$}{
    $triangle(v) \leftarrow$ \# of triangles that contain $v$\;
  }
  $k \leftarrow0$\;

  \While{$V\neq\emptyset$}{
  $k \leftarrow k+1$\;\\
  $unprocessedVertices = copy(V)$\;

  \While{$unprocessedVertices\neq\emptyset$}{
  Vertex $v = pop(unprocessedVertices)$\;

  \If{$triangle(v)<k$}{
    \ForAll{triangle $v$ of $t$}
    {
      \For{both other vertices $w$ of $t$}
      {
        $triangle(w)\leftarrow triangle(w)-1$\;\\
        $unprocessedVertices.add(w)$\;
      }
    }
    $v$.coreNumber$ \leftarrow k-1$\;\\
    remove $v$ from $V$\;
  }
  }
  }
\caption{Vertex triangle $k$-core decomposition algorithm}
\end{algorithm}

\textbf{Initialization.}\emph{ }The calculation of \emph{triangle($v$)}
is quadratic for a vertex $v$, so globally cubic complexity.

\textbf{Copied in} \textbf{\emph{unprocessedVertices}}\textbf{.} The
$k$ number goes from $0$ to $\frac{(n-1)(n-2)}{2}$. There are therefore
in total no more than $n^{3}$ edges copied in \emph{unprocessedVertices}.

\textbf{Unprocessed after a remove.} When a vertex is removed, all
other vertices which formed a triangle with the removed vertex are
added to \emph{unprocessedVertices}. There are therefore in total
no more than $n^{3}$ vertices added to \emph{unprocessedVertices}
after a remove.

\textbf{\emph{$\Rightarrow$ }}\textbf{Global complexity}\textbf{\emph{:}}\textbf{
}$\mathcal{O}(n^{3})$

\section{Experimental Evaluation}

How to compare two graph decompositions? What do we measure? How can
we assert that a decomposition is better than another one? Once we
have decided our comparison criteria, on which and on how many graphs
do we run tests? In this part, we propose to answer this questions
according to certain criteria we chose. This doesn't necessarily mean
that we got the best and unique answer --which we think doesn't actually
exists-- but that we analysed the problem within a perspective.

What should a \textquotedbl{}good\textquotedbl{} graph decomposition
look like? If you consider the core with the highest $k$, that we
will call the \emph{best community}, we would expect it to be limited
in terms of vertices. The cohesiveness of the best community should
be close to the one of the clique composed by the same vertices. But
finding a good community is not all, the graph decomposition should
list wider communities --not as good and small as the best one--,
but still interesting enough for being analysed. The speed and the
regularity of the increase (resp. decrease) of the number of vertices
while decreasing (resp. increasing) the level number, is fundamental
for our analysis.

In our study, we focused on these two aspects : the best community
and the core size decrease. A few examples of graph decompositions
is enough to notice the most important point of this whole study.
No graph decomposition --of the three we are trying to compare-- is
better than the other in any case. We cannot assert anything in absolute
terms about the best community size, neither can we about the core
size decrease. For a big set of graphs we calculate several individual
measures. Then average values of these metrics are analysed and compared.
The measures we made are detailed in the corresponding paragraphs.
First we precise on which graphs we run our decompositions in order
to compare them.

\subsection{Graph samples\label{sub:Measures}}

\textbf{Exhaustive Graph.} The first possibility is to generate all
graphs with $n$ vertices. There are $2^{\binom{n}{2}}=2^{n(n-1)/2}$
graphs with n vertices. We made it for $5$ vertices --$1024$ graphs--
and for $6$ vertices --$32,768$ graphs-- but for $7$ vertices --$2,097,152$
graphs-- it exceeds the capacity of our software.

\textbf{Random Graph.} To avoid memory issues, we decided to test
on random graph. Given a number of vertices $n$, there are $\binom{n}{2}=\frac{n(n-1)}{2}$
possible edges. We add each edge to the graph with a probability of
$1/2$. We construct $10,000$ graphs this way. We did the study for
many different $n$, but we kept the results for $9$, $15$, $25$
and $45$ vertices.

\textbf{Real Graph.} Real graphs are as \textquotedbl{}random\textquotedbl{}
as those we built. For example, in social network there is a high
probability of having the friend of someone's friend be their friend.
In other words, there is a high probability of a triangle. We have
tested our algorithms on several real graphs. We had some memory issues,
due to the kind of Java objects we use in our algorithms. The biggest
real graph we managed to decompose in the three shown ways, and the
only one we analysed, is called ca-HepPh which can found in \cite{key-8}.
It is the graph of Collaboration network of arXiv in High Energy Physics,
it contains $12,008$ nodes and $118,489$ edges.

\subsection{Best community}

In order to discuss about graph decomposition algorithm efficiency
for finding the best community, we define two kinds of accuracy measures.
\begin{defn}
We define the \emph{best level size} as the number of vertices with
the highest core number. \end{defn}
\begin{rem}
For the three decompositions of the preceding example $G_{0}$, the
best level size is $5$. \end{rem}
\begin{defn}
We define\emph{ best level clique-density} as the number of edges
divided by the number of possible edges in the best community, i.e.
$\frac{n}{\binom{n-1}{2}}$.\end{defn}
\begin{example}
In the example $G_{0}$, the best community is a clique, so the density
is $1$.
\end{example}
\textbf{Expected.} A \textquotedbl{}good\textquotedbl{} graph decomposition
should have a little best level size mean and higher best level density.
It is obvious that both ideas are equivalent. The best communities
given by different decomposition are nested. Thus the density decreases
as the best community size increases. It has to be mentioned that
the classical density, i.e. the number of vertices divided by the
number of edges, is not relevant for this study, because with this
definition a clique could have a lower density than a subgraph in
which it is included, which is a situation we do not want.

\begin{figure}[h]
\begin{centering}
\includegraphics{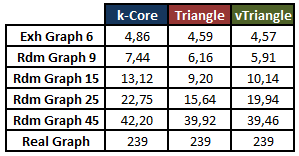}
\par\end{centering}

\protect\caption{Best level size comparison\label{fig:Best-level-size}}
\end{figure}

\begin{figure}[h]
\begin{centering}
\includegraphics{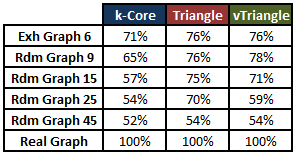}
\par\end{centering}

\protect\caption{Best level clique-density comparison\label{fig:Best-level-clique-density}}
\end{figure}

\textbf{Results.} As seen in Figure \ref{fig:Best-level-size}, the
vertex triangle decomposition seems to be the decomposition that finds
the shortest community, on average, in very small graphs; triangle
$k$-core in wider but still small graphs; finally vertex triangle
again. The $k$-core decomposition finds bigger communities in small
graphs but the gap is not as significant when the size increases.
Unfortunately, on all tests we have done on real graphs, the best
community was the same with the three decompositions, and was always
a clique. We tried to increase the number of vertices in random tests,
but the hierarchy on finding the best community on average was always
the same: first the vertex triangle, then the triangle $k$-core,
and then $k$-core at the end. As for the clique-density (Figure \ref{fig:Best-level-clique-density}),
it is higher when the community size is smaller, which is the case
for the vertex triangle decomposition.

Regarding the search of the best communities, both triangle decompositions
are better than $k$-core. But it is still questionable to assert
that vertex triangle $k$-core is better than (edge)-triangle $k$-core.

\subsection{Core size decrease}

After considering the best community, we should observe the rest of
the vertices.
\begin{defn}
We call a \emph{core} the set of vertices that, given an integer $k$,
have $\mbox{core number}\leq k$.
\end{defn}
In how many and which levels are held the remaining vertices? Are
they well distributed into all $k$-levels or abruptly bonded together
in only few levels? Which graph decomposition enables the most regular
core size decrease? We decided to proceed with the following measures
:

\textbf{Highest core number.} In the preceding example it is $4$
for $k$-core (Figure \ref{fig:-core-decomposition-of}), $3$ for
triangle $k$-core (Figure \ref{fig:-truss-decomposition-of}) and
$6$ for vertex triangle (Figure \ref{fig:Vertex-triangle--core}).
\begin{defn}
We define the \emph{level number }as\textbf{ }the quantity of not
empty levels. \end{defn}
\begin{rem}
In the example for $G_{0}$ it is 3 for $k$-core, $4$ for triangle
$k$-core and $5$ for vertex triangle. It can be seen graphically
as the number of different colours. \end{rem}
\begin{defn}
We define the \emph{Root Mean Square (RMS)}\textbf{ }as the quadratic
mean of the distribution of the vertices in the different levels
\begin{align*}
\sqrt{\sum_{k}\frac{n(k)^{2}}{n^{2}}}
\end{align*}
where $n(k)$ is the number of vertices in the $k$-core subgraph
and $n$ is the total number of vertices. \end{defn}
\begin{rem}
In the example for $G_{0}$ it is $0.63$ for $k$-core, $0.61$ for
triangle $k$-core and $0.59$ for vertex triangle.
\begin{align*}
\end{align*}

\end{rem}
\textbf{Expected.} The highest core number is not an effective measure.
The reason is that the vertex triangle decomposition lets a lot of
empty levels. If you consider only not empty levels, the core number
for these levels rises quadratically. However, the level number tells
about the possibility of vertex segmentation. More levels means more
different ways to choose the wideness of communities. It is quite
useless to have very small levels or very big levels, in terms of
size. Ideally, the levels have all the same size. On the one hand,
the RMS is minimal when it is the case. On the other hand, the RMS
is lower when the number of not empty levels increases. For these
two reasons we calculate RMS, and we expected it to be low for \textquotedbl{}good\textquotedbl{}
decomposition.

\begin{figure}[h]
\begin{centering}
\includegraphics{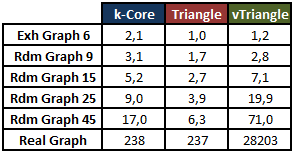}
\par\end{centering}

\protect\caption{Highest core number comparison\label{fig:Highest-core-number}}
\end{figure}

\begin{figure}[h]
\begin{centering}
\includegraphics{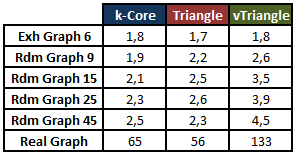}
\par\end{centering}

\protect\caption{Level number comparison\label{fig:Level-number-comparison}}
\end{figure}

\begin{figure}[h]
\begin{centering}
\includegraphics{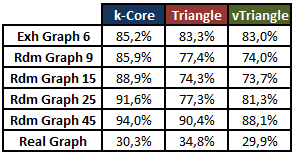}
\par\end{centering}

\protect\caption{Root Mean Square comparison\label{fig:Root-Mean-Square}}
\end{figure}

\textbf{Results.} As provided, the highest core number, as seen in
Figure \ref{fig:Highest-core-number}, is bigger on average in the
vertex triangle decomposition than the other two, because the best
core number evolution is quadratic for it and linear for the other
two. What was not obvious is that the highest core number for triangle
$k$-core is significantly lower than the $k$-core. As for the level
number, as seen in Figure \ref{fig:Level-number-comparison}, it is
clearly the vertex triangle decomposition which has the best. If the
triangle $k$-core is better in that way than the $k$-core in small
graphs, the real graph confirms that the tendency goes on the other
way while graph size increases. Triangle $k$-core decomposition offers
far less different levels than the $k$-core decomposition.

Our study has not made it clear about the root mean square. Considering
the random tests, it appears that both triangle decompositions are
fighting for the first place for RMS, as seen in Figure \ref{fig:Root-Mean-Square},
with a slight advantage for vertex triangle. This tendency is confirmed
when random tests are done with more than $45$ vertices. The real
graph shows it differently. Vertex triangle is still leading, but
challenged by $k$-core, and triangle $k$-core RMS values are far
bigger. Given these results we ran test on other real graphs to check
that it was not an isolated characteristic of the chosen graph. Maybe
the fact that triangle $k$-core decomposition has less levels, explains
its bigger RMS.

\begin{figure}[h]
\begin{centering}
\includegraphics[scale=0.85]{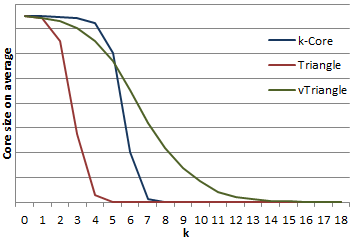}
\par\end{centering}

\protect\caption{Core size decrease for random $15$ vertices graph\label{fig:Core-size-decrease}}

\end{figure}

\begin{figure}[h]
\begin{centering}
\includegraphics[scale=0.9]{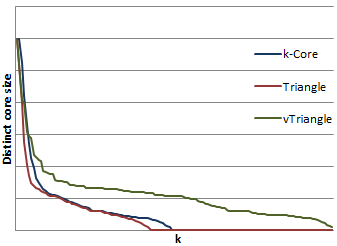}
\par\end{centering}

\protect\caption{Distinct core size decrease for arXiv graph\label{fig:Core-size-decrease- arxiv}}

\end{figure}

Figure \ref{fig:Core-size-decrease} show the decrease of core size
while $k$ increases. For this last, values are considered as the
average over $10,000$ graphs. Figure \ref{fig:Core-size-decrease- arxiv}
show as well the core size decrease as $k$ increases, for the real
arXiv graph, with the difference that we don't considerate $k$ if
there are no vertices with exactly $k$ as their core number.

Considering these curves of core size decrease, the vertex triangle
decomposition (green curve) seems to have the slowest and most regular
decrease, on both random and real graphs. As for $k$-core and triangle
$k$-core, it is difficult to tell. The core size for $k$-core is
bigger for little $k$ but decreases abruptly. It is more useful to
decrease slowly for high $k$, in order to distinguish good and very
good communities rather than bad and very bad ones.

Regarding core size decrease, the best decomposition is the vertex
triangle decomposition in various aspects. It has more levels, quite
regularly filled, which enables to select a community with a more
precise size. The $k$-core and triangle $k$-core are both bad at
it, but the triangle $k$-core is more regular in its core size decrease.

\subsection{Algorithm execution time}

\begin{figure}[h]
\begin{centering}
  \includegraphics[scale=0.9]{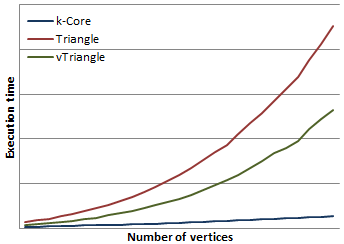}
\par\end{centering}

\protect\caption{Execution time for the three algorithms\label{fig:Execution-time-of}}
\end{figure}

In order to verify the complexity of the three algorithms, we ran
them on random graphs with different number of vertices. The way of
constructing random graph is the same as described in \ref{sub:Measures}.
For a number of vertices, the program creates randomly $1000$ graphs
and the three algorithms calculate the decompositions. The mean is
done on the $1000$ execution time. These means established for number
of vertices from $10$ to $40$, are reported in Figure \ref{fig:Execution-time-of}.

\begin{figure}[h]
\begin{centering}
\includegraphics[scale=0.9]{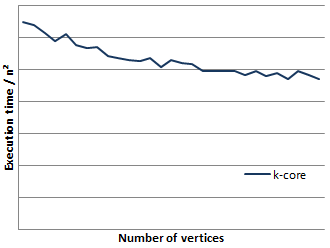}
\par\end{centering}

\protect\caption{Quadratic Verification\label{fig:Quadratic-Verification}}

\end{figure}

\begin{figure}[h]
\begin{centering}
\includegraphics[scale=0.9]{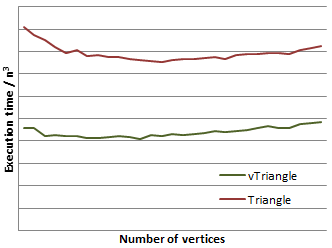}
\par\end{centering}

\protect\caption{Cubic Verification\label{fig:Cubic-Verification}}

\end{figure}

As expected, both triangle decompositions take far more time than
$k$-core. To check whether the complexity is cubic for triangle decompositions
and quadratic for $k$-core decomposition, the figures \ref{fig:Quadratic-Verification}
and \ref{fig:Cubic-Verification} show the evolution of execution
times of algorithms divided by their complexity ($\mathcal{O}(n^{2})$
or $\mathcal{O}(n^{3})$).

The k-core decomposition algorithm is definitely quadratic, but both
triangle algorithms we used are a little more than cubic. It can be
explained by the fact we used \emph{ArrayList} Java objects for a
more comfortable implementation, whose basic operations are not executed
in constant time.

\section{Conclusions}

We have shown that $k$-core decomposition are efficient but not very
precise: it can be computed in a quadratic time --versus cubic time
for the others decompositions-- but the core size decrease is not
very regular. Sacrificing complexity for results, we got a far more
useful decomposition using the novel vertex triangle $k$-core, better
than the (edge) triangle $k$-core: core size decrease for vertex
triangle $k$-core is more regular and it converges to a smaller community.
In a future, some new $p$-functions for generalized core decomposition
could perhaps give us better results.

\section*{Acknowledgment}
\addcontentsline{toc}{section}{Acknowledgment}
We thank our professor Michalis Vazirgiannis, and his postdoctoral researcher
Fragkiskos Malliaros, who proposed the subject and provided insight and
expertise that greatly assisted the research, although they may not agree with
all of the interpretations of this paper.

\end{document}